\documentclass[letterpaper,english,aps]{revtex4}
\usepackage{mathptmx}
\usepackage[T1]{fontenc}
\usepackage{amstext}
\usepackage{graphicx}
\usepackage{amssymb}

\makeatletter
\@ifundefined{textcolor}{}
{%
 \definecolor{BLACK}{gray}{0}
 \definecolor{WHITE}{gray}{1}
 \definecolor{RED}{rgb}{1,0,0}
 \definecolor{GREEN}{rgb}{0,1,0}
 \definecolor{BLUE}{rgb}{0,0,1}
 \definecolor{CYAN}{cmyk}{1,0,0,0}
 \definecolor{MAGENTA}{cmyk}{0,1,0,0}
 \definecolor{YELLOW}{cmyk}{0,0,1,0}
 }

\@ifundefined{showcaptionsetup}{}{%
 \PassOptionsToPackage{caption=false}{subfig}}
\usepackage{subfig}
\makeatother

\usepackage{babel}

\begin{document}

\title{Thermodynamic approach to the dewetting instability in ultrathin
films}

\author{$^{\text{1}}$N. Shirato,$^{\text{2}}$H. Krishna, and $^{\text{1,3,4}}$R.
Kalyanaraman%
\thanks{All correspondence to be addressed to ramki@utk.edu%
}}

\affiliation{$^{\text{1}}$Department of Materials Science and Engineering, University
of Tennessee, Knoxville, TN 37996}

\affiliation{$^{2}$Department of Physics, Washington University in St. Louis,
MO 63130}

\affiliation{$^{\text{3}}$Department of Chemical and Biomolecular Engineering,
University of Tennessee, Knoxville, TN 37996}

\affiliation{$^{\text{4}}$Sustainable Energy Education and Research Center, University
of Tennessee, Knoxville, TN 37996\thispagestyle{empty}}
\begin{abstract}
The fluid dynamics of the classical dewetting instability in ultrathin
films is a non-linear process. However, the physical manifestation
of the instability in terms of characteristic length and time scales
can be described by a linearized form of the initial conditions of
the film's dynamics. Alternately, the thermodynamic approach based
on equating the rate of free energy decrease to the viscous dissipation
{[}de Gennes, C. R. Acad. Paris. v298, 1984{]} can give similar information.
Here we have evaluated dewetting in the presence of thermocapillary
forces arising from a film-thickness (h) dependent temperature. Such
a situation can be found during pulsed laser melting of ultrathin
metal films where nanoscale effects lead to a local h-dependent temperature.
The thermodynamic approach provides an analytical description of this
thermocapillary dewetting. The results of this approach agree with
those from linear theory and experimental observations provided the
minimum value of viscous dissipation is equated to the rate of free
energy decrease. The flow boundary condition that produces this minimum
viscous dissipation is when the film-substrate tangential stress is
zero. The physical implication of this finding is that the spontaneous
dewetting instability follows the path of minimum rate of energy loss.
\end{abstract}
\maketitle

\section{Introduction}

Investigations of thin film morphology evolution and control is of
fundamental and technological interest. In particular, spontaneous
self-organizing processes \cite{Nicolas77} that lead to nanostructure
formation in a reliable way have attracted tremendous attention. The
resulting nanostructures can have novel behavior as well as be utilized
in a wide variety of technologies, such as, energy harvesting \cite{stuart98,pillai06,ColeAPL06},
biomedicine\cite{NamScience03,DanielChemRev04}, spintronics \cite{ParkJAP02},
photonics \cite{ShipwayChemPhysChem00} and magnetism \cite{sun00,krishna08}.
One example of a self-organizing process is the spontaneous dewetting
of a continuous liquid film from a surface. A scientific understanding
of dewetting has implications to many industrial applications, including
in the deicing of airplane wings with non-wettable surfaces, in preventing
hydroplaning of automobiles on wet roads due to thin continuous layer
of water and in designing chemicals to prevent the break-up of the
lachrymal film that protects the cornea of the eye. Another growing
application of dewetting is in the fabrication of nanoscale structures
in a robust, controllable and cost-effective manner. The extensive
studies of polymer thin films and growing number of investigations
of dewetting in metallic thin films is indicative of this technological
interest and also to the need for a deeper understanding of the phenomenon
\cite{Reiter92,Seemann01,redon91,Herminghaus98-a,stange97-b,thiele01,henley05,favazza06d,Trice06a,Kondic09}. 

The classical dewetting instability in thin films can be interpreted
as a competition between two energy terms. For the case of a large
number of polymer or metallic films studied, these two energies correspond
to the surface tension and the attractive intermolecular dispersion
force between the film-substrate and film-vacuum interfaces mediated
by the film material. As shown first by Vrij \cite{vrij66,vrij68},
the instability can be described from an energetic viewpoint by evaluating
the thermodynamic free energy change of the system under perturbations
to the film height. The prediction from such an energy analysis is
that for certain perturbation wave vectors, the film enters an unstable
state and thus, can spontaneously dewet. As a result, studies of dewetting
have focused largely on the fluid dynamics of the film, through which
it is possible to obtain the relationship between the rate of growth
or decay of surface perturbations to their wave vector, i.e. the dispersion
relation. However, the fluid dynamics for even the simplest dewetting
scenario, such as the example above, is a highly nonlinear process,
and, while addressable by many numerical techniques \cite{Sharma,becker03,trice07b,Khenner09},
is often evaluated through a linear analysis in order to achieve physical
insights into dewetting. An alternate approach to quantitatively evaluate
dewetting is thermodynamics. Fluid flow pathways can be analyzed through
thermodynamic considerations in which the conversion of useful internal
energy to external energy loss via heat, such as by viscous dissipation,
is used to quantify the behavior \cite{degennes84,deGennes85}. 

In this work, we show that such an approach can provide meaningful
insight into the nature of fluid flow as well as the energy pathway
for dewetting instabilities. Specifically, we have applied the thermodynamic
formulation to the case of dewetting in which film thickness dependent
Marangoni or thermocapillary forces are also present. Such a situation
has been observed in the melting of thin metallic films by nanosecond
pulsed lasers \cite{trice08,KrishnaPCCP09}. In our thermodynamic
analysis the rate of thermodynamic free energy decrease due to film
thickness fluctuations is balanced with the rate of energy loss due
to viscous flow, i.e. viscous dissipation. This leads to a analytical
description of the dewetting process without explicit need to solve
the the height evolution dynamical equation from the Navier-Stokes
(NS) equation. The thermodynamic and linear approach show identical
results for classical dewetting. For thermocapillary dewetting, the
two approaches agree only if the minimum viscous dissipation is evaluated.
This minimum was found to occur for a particular flow boundary condition,
which related the pressure gradient with the thermocapillary forces,
and resulted in zero tangential stress at the film-substrate interface.
Besides this physical insight into the fluid flow, the thermodynamic
analysis also showed that the dewetting pathway is one in which the
rate of energy loss is minimized.

\section{Theory}

For completeness, we first begin by summarizing the derivation of
the thin film fluid velocity for a one-dimensional (1D) incompressible
fluid from the NS equation within the lubrication approximation. A
complete analysis is provided in ref. \cite{Kondic03}. In this approximation
the average or unperturbed thickness $h_{o}$ of the film is much
smaller than the in-plane dimension (x), as a result of which, the
only velocity change of importance occurs along the thickness or z-direction.
Furthermore, because of the small thickness of the film, inertial
effects can be neglected and so the flow is dominated by the viscous
effects. Using the above approximations, the NS equation for the steady-state
condition ($\rho$$\frac{dv}{dt}=0)$ in the $x$-direction is given
by:\begin{equation}
\nabla P=\eta\frac{d^{2}v}{dz^{2}}\label{eq:NS equation}\end{equation}

\noindent where $v$ is the $x$-component of the liquid velocity,
$\nabla P=\frac{dp}{dx}$ is the pressure gradient in the direction
of flow $x$, and $\eta$ is the dynamic viscosity. By integrating
the velocity $v$ as a function of height z we get:\begin{equation}
v=\nabla P\frac{z^{2}}{2\eta}+az+b\label{eq:velocity-x-direction}\end{equation}
Typical boundary conditions used to analyze the classical dewetting
instability are the no-slip condition at the film-substrate, so $v(h=0)=0$,
and a stress-free boundary condition at the top film surface. The
no-slip condition results in $b=0$. At the top surface, we introduce
the h-dependent Marangoni effect by equating the shear stress to the
surface tension gradient: $\eta\frac{dv}{dz}|_{h_{0}}=\frac{d\gamma(h)}{dh}\frac{dh}{dx}=\gamma_{h}\frac{dh}{dx}=(\gamma_{h}h'){}_{h_{o}}$,
where, $\gamma(h)$ is h-dependent surface tension of the film-vapor
interface, $\mid\gamma_{h}\mid=\mid\frac{d\gamma}{dh}\mid$ is the
magnitude of the height coefficient of surface tension, and $\frac{dh}{dx}=h'$
is the thickness/height gradient along the flow direction, with all
quantities evaluated at the average film thickness $h_{o}$. From
this, the velocity and velocity gradient in the $z$-direction can
be respectively expressed as: \begin{equation}
v=\frac{\nabla P}{2\eta}z^{2}-\frac{\nabla Ph_{0}-\gamma_{h}h'}{\eta}z\label{eq:Vx}\end{equation}

\noindent and \begin{equation}
\frac{dv}{dz}=\frac{\nabla P}{\eta}z-\frac{\nabla Ph_{0}-\gamma_{h}h'}{\eta}\label{eq:dVxoverdZ}\end{equation}
Based on this, one can now easily evaluate the rate of energy loss
due to viscous liquid flow, i.e. the viscous dissipation, per unit
volume $\dot{e}$ occurring in the film. This quantity is given by
\cite{guyon01}:\begin{equation}
\dot{e}=\eta(\frac{dv}{dz})^{2}=\frac{(\nabla P)^{2}}{\eta}z^{2}-\frac{2\nabla P(\nabla Ph_{0}-\gamma_{h}h')}{\eta}z+\frac{(\nabla Ph_{0}-\gamma_{h}h')^{2}}{\eta}\label{eq:Dissipation}\end{equation}

\noindent For the case of classical dewetting, i.e. in which Marangoni
forces are absent ($\gamma_{h}=0$), the viscous dissipation will
be: \begin{equation}
\dot{e}_{C}=\frac{(\nabla P)^{2}}{\eta}(z-h_{o})^{2}\label{eq:Dissipation-noMarangoni}\end{equation}

\noindent where the superscript denotes classical. 

\noindent Next, we can evaluate the rate of thermodynamic free energy
change $\dot{\Delta F}$ for fluctuations/perturbations to the initial
height of the film. Since we are primarily concerned with the dewetting
instability, we will use the classical approach proposed by Vrij where-in
the film-vapor surface tension energy competes with the attractive
dispersion energy. Film height perturbations will increase the top
film surface area and so surface tension increases the overall thermodynamic
free energy of the film. On the other hand, the long range attractive
dispersion energy varies as $\Pi^{''}=A/2\pi h^{4}$, where A is the
Hamaker coefficient with negative sign, leading to an overall decrease
in thermodynamic free energy. As Vrij showed, it is the free energy
decrease resulting from competition between these two energy terms
that drives the dewetting instability. Here, we evaluate the rate
of this free energy change by expressing the height perturbations
as Fourier components of type: \begin{equation}
h(x,t)=h_{0}+\epsilon e^{\sigma t}e^{-ikx}\label{eq:hxt}\end{equation}
where the perturbation has an amplitude of $\epsilon$, a characteristic
temporal decay rate $\sigma$ and a corresponding wave vector $k$.
In this work, we explore the thermodynamic solution for a thermocapillary/Marangoni
problem where the surface tension is dependent on the local height
but independent of the $x$-position on the surface. Consequently,
the surface tension change does not contribute to a change in free
energy. This can be shown as follows. The standard procedure is to
calculate the rate of total free energy change due to the height perturbation
and evaluate it over one wavelength of the perturbation \cite{vrij66}.
In this case, the resulting rate of energy change will be

\begin{equation}
\Delta F_{single\, wavelength}=\int_{0}^{\lambda}(\frac{1}{2}\gamma\mid\frac{\partial h}{\partial x}\mid^{2}+\frac{1}{2}\Pi^{''}\Delta h{}^{2}+\frac{1}{2}\gamma_{h}\Delta h)dx\label{eq:DelF}\end{equation}

One can immediately see that when the magnitude of the surface tension
derivative with respect to height ($\gamma_{h}=d\gamma/dh$) is independent
of the $x$-position along the film, the integral of the third term
is $\int_{0}^{\lambda}\frac{1}{2}\frac{d\gamma}{dh}\epsilon e^{-ikx+\sigma t}dx=0$.
Consequently, to calculate the rate of change in the free energy,
we can ignore the contribution from $\gamma_{h}$, and express it
as the difference due to the initial and perturbed film thickness
at any position $x$ \cite{vrij68} as:

\noindent \begin{equation}
\dot{\Delta F}=\frac{\partial}{\partial t}\left[\frac{1}{2}\gamma\mid\frac{\partial h}{\partial x}\mid^{2}+\frac{1}{2}\Pi^{''}\Delta h{}^{2}\right]=\sigma(\gamma k^{2}+\frac{A}{2\pi h_{0}^{4}})\epsilon^{2}e^{2\left(\sigma t-ikx\right)}\label{eq:dFoverdt}\end{equation}
 The first term in the expression on the right hand side is the rate
of increase of surface tension energy, and the second one is the rate
of change in energy from the dispersive interaction.

\section{Results }

\subsection{\noindent Dispersion relation for classical dewetting}

\noindent Here we compare the characteristic dewetting length scales
obtained from fluid dynamics versus the thermodynamic approach. The
typical approach to obtain the dispersion relation between the rate
$\sigma$ and wave vector $k$ has been to describe the rate of change
in film height based on the NS equation and mass conservation \cite{vrij66,vrij68}.
For the classical dewetting instability, i.e. without Marangoni effects,
the resulting dynamics is described by the equation \cite{sharma86,favazza06d}: 

\begin{equation}
3\eta\frac{\partial h}{\partial t}=-\bigtriangledown.\left(\gamma h{}^{3}\bigtriangledown.\bigtriangledown^{2}h-\frac{A}{2\pi h}\bigtriangledown h\right)\label{eq:Dynamics-noMarangoni}\end{equation}

As is evident, this equation is non-linear in $h$ and presents considerable
challenges towards achieving an analytical description of dewetting
that could provide simple but physically insightful information about
the instability. Consequently, a prevalent approach is a solution
afforded by linear stability analysis. Hence, after an expansion of
the right hand side of Eq. \ref{eq:Dynamics-noMarangoni}:\begin{eqnarray}
3\eta\frac{\partial h}{\partial t} & = & (-k^{2}\epsilon e^{\sigma t-ikx})\left(\gamma k^{2}(h_{o}^{3}+6h_{o}^{2}\epsilon e^{\sigma t-ikx}+9h_{o}\epsilon^{2}e^{2(\sigma t-ikx)}+4\epsilon^{3}e^{3(\sigma t-ikx)})+\frac{Ah_{o}}{2\pi(h_{o}+\epsilon e^{\sigma t-ikx})^{2}}\right)\label{eq:Dynamics-noMarangoniExpand}\end{eqnarray}
and applying by keeping terms only linear in the perturbation amplitude
$\epsilon$, as required by LSA, Eq. \ref{eq:Dynamics-noMarangoniExpand}
reduces to :\begin{eqnarray}
3\eta(\epsilon\sigma e^{\sigma t-ikx}) & = & (-k^{2}\epsilon e^{\sigma t-ikx})\left(\gamma k^{2}h_{o}^{3}+\frac{A}{2\pi h_{o}}\right)\label{eq:Dynamics-noMarangoniLinear}\end{eqnarray}

Consequently, the resulting dispersion relation is given by \cite{trice08}:\begin{equation}
\sigma_{C}^{LSA}=-\frac{h_{0}^{3}k^{2}}{3\eta}(\gamma k^{2}+\frac{A}{2\pi h_{0}^{4}})\label{eq:sigma-Classical-LSA}\end{equation}

The characteristic (or classical) dewetting length scale $\Lambda_{C}^{LSA}$
can be obtained from the dispersion relation by the maxima condition
$\frac{d\sigma}{dk}=0$ and leads to:\begin{equation}
\Lambda_{C}^{LSA}=\left(-\frac{8\pi^{2}\gamma h_{o}^{3}}{\frac{A}{2\pi h_{0}}}\right)^{1/2}=\sqrt{-\frac{16\pi^{3}\gamma}{A}}h_{o}^{2}\label{eq:lambda-classical-LSA}\end{equation}

\noindent where the superscript refers to LSA. 

On the other hand, the thermodynamic (TH) approach is based on equating
the rate of free energy change (Eq. \ref{eq:dFoverdt}) to the total
viscous dissipation in the film. We can calculate the total viscous
dissipation per unit area $\dot{E}$ for the liquid film by integrating
over the film thickness as follows:

\noindent \begin{equation}
\dot{E}_{C}=\intop_{0}^{h_{0}}\dot{e_{C}}dz=\frac{(\nabla P)^{2}}{3\eta}h_{0}^{3}\label{eq:Evtotal_Class}\end{equation}

\noindent The next, and important, step in evaluating this integral
is to relate the pressure gradient to the film height through a volume
conservation argument. Volume conservation requires that the rate
of change of film height $\partial h/\partial t$ be related to the
flux of liquid flow $J(x)$ as $\partial h/\partial t=-\nabla\centerdot J\left(x\right)$.
To evaluate this we have used the thin film lubrication approximation
in which flux can be written in terms of the height-averaged liquid
velocity $<v>$ as \cite{Kondic03}:

\noindent \begin{equation}
J\left(x\right)=h_{o}<v>=h_{o}.(\frac{1}{h_{o}}\int_{0}^{h_{o}}vdz)=-\frac{\nabla P}{3\eta}h_{o}^{3}\label{eq:Flux}\end{equation}

\noindent from which we can express the volume conservation equation
as:\begin{equation}
\frac{\partial h}{\partial t}=-\nabla.J=\frac{\nabla^{2}P}{3\eta}h_{o}^{3}\label{eq:volume conserve}\end{equation}
By rearranging terms we get the desired relation for the pressure
gradient as follows:

\noindent \begin{equation}
\nabla P=\int\nabla^{2}Pdx=\frac{i}{k}(\frac{3\eta\sigma}{h_{o}^{3}})\epsilon e^{\sigma t-ikx}\label{eq:nabla P}\end{equation}
where we have made use of Eq. \ref{eq:hxt}. Using this expression
in Eq. \ref{eq:Evtotal_Class}, the total dissipation is:

\noindent \begin{equation}
\dot{E_{C}}=-\frac{3\eta}{h_{0}^{3}k^{2}}\sigma^{2}\epsilon^{2}e^{2(\sigma t-ikx)}\label{eq:Evtotalsub_Class}\end{equation}

\noindent Then, equating Eq. \ref{eq:Evtotalsub_Class} with the rate
of decrease of free energy, Eq. \ref{eq:dFoverdt}, $\dot{\Delta F}=\dot{E_{C}}$,
and expressing the result in terms of $\sigma$ we get the dispersion
relation from the thermodynamic approach:

\begin{equation}
\sigma_{C}^{TH}=-\frac{h_{0}^{3}k^{2}}{3\eta}(\gamma k^{2}+\frac{A}{2\pi h_{0}^{4}})\label{eq:sigma-classical-TH}\end{equation}

\noindent from which, the classical dewetting length scale $\Lambda_{C}^{TH}$
can be expressed as:\begin{equation}
\Lambda_{C}^{TH}=\left(-\frac{8\pi^{2}\gamma h_{_{O}}^{3}}{\frac{A}{2\pi h_{o}}}\right)^{1/2}=\sqrt{-\frac{16\pi^{3}\gamma}{A}}h_{o}^{2}\label{eq:lambda-classical-TH}\end{equation}

\noindent where the superscript and subscript refers to thermodynamic.
As expected, the fluid dynamics described by LSA and the TH approach
give identical results for the classical dewetting instability \cite{degennes03-a,degennes84}.
On the other hand, as we show next, dewetting with Marangoni (or Thermocapillary)
forces requires a more stringent evaluation of the viscous dissipation
in order to provide results comparable to LSA.

\subsection{\noindent Dispersion relation for thermocapillary dewetting from
thermodynamics }

\noindent As in the previous section, we analyze the LSA and TH approaches
for dewetting in the presence of thickness-dependent Marangoni effects.
In order to relate our work to experimental observations, we describe
LSA results for the case when ultrathin metal films on SiO$_{\text{2}}$
substrates are melted by nanosecond ultraviolet wavelength laser pulses
\cite{Herminghaus98-a,henley05}. In this situation, there is a strong
thickness-dependent reflection and absorption of light by the thin
metal film which leads to a local h-dependent temperature of the liquid
metal film \cite{Trice06a}. In addition, as reported previously,
the temperature gradient along the plane of the film, $dT/dx$, generated
by this nanoscale heating effect, can have a positive or negative
sign depending upon the initial film thickness $h_{o}$ \cite{trice08}.
With this, the boundary condition describing the $h$-dependent Marangoni
effect can be rewritten in the form of a thermocapillary effect as
follows: \begin{equation}
\eta\frac{dv}{dz}|_{h_{0}}=\frac{d\gamma(h)}{dh}\frac{dh}{dx}=-\mid\gamma_{T}\mid T_{h}h'\label{eq:shear stress}\end{equation}

Where $\gamma_{T}=\frac{d\gamma}{dT}$ is the temperature coefficient
of surface tension, $T_{h}=dT/dh$ is the film height-dependent temperature
and $h'=dh/dx$. Given that all metals have a negative value of $\gamma_{T}$,
we have expressed the boundary condition in a more conventional form
using $\gamma_{T}=-\mid\gamma_{T}\mid$. In this scenario, the resulting
dynamical equation of the film height is given by:

\begin{equation}
3\eta\frac{\partial h}{\partial t}=-\bigtriangledown.\left(\gamma h{}^{3}\bigtriangledown.\bigtriangledown^{2}h-\frac{A}{2\pi h}\bigtriangledown h+\frac{3}{2}h^{2}\nabla\gamma\right)\label{eq:dynamical equation}\end{equation}

Applying steps similar to to the classical case, the right hand side
of Eq. \ref{eq:dynamical equation} is:

\begin{eqnarray}
3\eta\frac{\partial h}{\partial t} & = & -(\epsilon e^{\sigma t-ikx})\left(\gamma k^{4}(h_{o}+\epsilon e^{\sigma t-ikx})^{2}\left(h_{o}+4\epsilon e^{\sigma t-ikx}\right)\right.\label{eq:dynamical eqExpanded}\\
 &  & \left.+\frac{A(k^{2}h_{o})}{2\pi(h_{o}+\epsilon e^{\sigma t-ikx})^{2}}+3k^{2}\mid\gamma_{T}\mid T_{h}(h_{o}+\epsilon e^{\sigma t-ikx})\left(\frac{h_{o}}{2}+\frac{3}{2}\epsilon e^{\sigma t-ikx}\right)\right)\end{eqnarray}

which, upon applying LSA, reduces to:

\begin{eqnarray*}
3\eta(\epsilon\sigma e^{\sigma t-ikx}) & = & -(\epsilon e^{\sigma t-ikx})\left(\gamma k^{4}h_{o}^{3}+\frac{Ak^{2}}{2\pi h_{o}}+\frac{3}{2}h_{o}^{2}k^{2}\mid\gamma_{T}\mid T_{h}\right)\end{eqnarray*}

\noindent and the resulting dispersion relation is\cite{trice08}:\begin{equation}
\sigma_{TC}^{LSA}=-\frac{h_{0}^{3}k^{2}}{3\eta}(\gamma k^{2}+\frac{A}{2\pi h_{0}^{4}}+\frac{3}{2}\frac{\mid\gamma_{T}\mid T_{h}}{h_{0}})\label{eq:LSA-Disp-TC}\end{equation}

\noindent From Eq. \ref{eq:LSA-Disp-TC} the characteristic dewetting
length scale in the presence of thermocapillary effects, $\Lambda_{TC}^{LSA}$,
can be expressed as:\begin{equation}
\Lambda_{TC}^{LSA}=\frac{2\pi}{k}=\left(-\frac{16\pi^{3}\gamma}{A+3\pi\mid\gamma_{T}\mid h_{o}^{3}T_{h}}\right)^{1/2}h_{o}^{2}\label{eq:LSA_LambdaTC}\end{equation}

\noindent where the superscript $TC$ denotes thermocapillary.

\noindent Next we evaluate the length scale using the TH approach
based on evaluating the the total viscous dissipation per unit area
$\dot{E}$ for the liquid film. First, the dissipation per unit volume
expressed in terms of the thermocapillary boundary condition is:\begin{equation}
\dot{e}_{TC}=\eta(\frac{dv}{dz})^{2}=\frac{\nabla P^{2}}{\eta}(z-h_{o})^{2}-2\nabla P\frac{\mid\gamma_{T}\mid T_{h}h'}{\eta}(z-h_{o})+\frac{(\mid\gamma_{T}\mid T_{h}h')^{2}}{\eta}\label{eq:ev_TC}\end{equation}

\noindent The total viscous dissipation per unit area $\dot{E}$ for
the liquid film is now:

\noindent \begin{eqnarray}
\dot{E}_{TC} & = & \intop_{0}^{h_{0}}\dot{e}_{TC}dz=\frac{\nabla P^{2}}{3\eta}h_{o}^{3}+\nabla P\frac{\mid\gamma_{T}\mid T_{h}h'}{\eta}h_{o}^{2}+\frac{(\mid\gamma_{T}\mid T_{h}h')^{2}}{\eta}h_{o}\label{eq:Evtotal_TC}\end{eqnarray}

\noindent Again, the next step in evaluating this integral is to relate
the pressure gradient to the film height through volume conservation
arguments, as done for the classical case.

\noindent \begin{equation}
J\left(x\right)=h_{o}<v>=h_{o}.(\frac{1}{h_{o}}\int_{0}^{h_{o}}vdz)=-\frac{\nabla P}{3\eta}h_{o}^{3}-\frac{\mid\gamma_{T}\mid T_{h}h'h_{o}}{2\eta}\label{eq:Flux-1}\end{equation}

\noindent from which we can express the volume conservation equation
as:\begin{equation}
\frac{\partial h}{\partial t}=-\nabla.J=\frac{\nabla^{2}P}{3\eta}h_{o}^{3}+\frac{\mid\gamma_{T}\mid T_{h}h''h_{o}^{2}}{2\eta}\label{eq:Vol conserv}\end{equation}

\noindent By rearranging terms and substituting the height perturbation,
Eq. \ref{eq:hxt}, into above we get:

\noindent \begin{equation}
\nabla^{2}P=\frac{3\eta}{h_{o}^{3}}\epsilon e^{\sigma t-ikx}+\frac{3\mid\gamma_{T}\mid T_{h}k^{2}}{2h_{o}}\epsilon e^{\sigma t-ikx}\label{eq:nablasqP}\end{equation}
The pressure gradient along the x-direction can now be obtained by
integrating Eq. \ref{eq:nablasqP} as follows: 

\noindent \begin{equation}
\nabla P=\int\nabla^{2}Pdx=\frac{i}{k}(\frac{3\eta\sigma}{h_{o}^{3}}+\frac{3\mid\gamma_{T}\mid T_{h}k^{2}}{2h_{o}})\epsilon e^{\sigma t-ikx}\label{eq:nablaPfinal_TC}\end{equation}
Then, on substituting the above relation for pressure gradient into
Eq. \ref{eq:Evtotal_TC}, we get the total viscous dissipation:

\noindent \begin{eqnarray}
\dot{E}_{TC} & = & \left\{ -\left(\frac{3\eta}{h_{0}^{3}k^{2}}\sigma^{2}+\frac{3\mid\gamma_{T}\mid T_{h}}{h_{0}}\sigma+\frac{3(\mid\gamma_{T}\mid T_{h})^{2}k^{2}h_{0}}{4\eta}\right)\right.\nonumber \\
 &  & \left.+(\frac{3\eta\mid\gamma_{T}\mid T_{h}}{h_{0}^{3}}\sigma+\frac{3(\mid\gamma_{T}\mid T_{h})^{2}k^{2}}{2h_{0}})\frac{h_{0}^{2}}{\eta}-\frac{h_{0}}{\eta}(\mid\gamma_{T}\mid T_{h})^{2}k^{2}\right\} \epsilon^{2}e^{2(\sigma t-ikx)}\label{eq:Evtotalsub}\end{eqnarray}

\noindent Finally, equating Eq. \ref{eq:Evtotalsub} with the rate
of decrease of free energy, Eq. \ref{eq:dFoverdt} and rearranging
the equation in terms of $\sigma$ we get an analytical dispersion
expression for thermocapillary dewetting as:

\begin{equation}
\sigma^{2}+\frac{h_{0}^{3}k^{2}}{3\eta}(\gamma k^{2}+\frac{A}{2\pi h_{0}^{4}})\sigma+(\frac{h_{0}^{3}k^{2}}{3\eta})\frac{(\mid\gamma_{T}\mid T_{h})^{2}k^{2}h_{0}}{4\eta}=0\label{eq:Thermo-Disp-TC}\end{equation}
It is important to enote that Eq. \ref{eq:Thermo-Disp-TC}, which
comes from the thermodynamic approach, differs from the LSA result,
Eq. \ref{eq:LSA-Disp-TC}. One can note that there is substantial
difference between the two solutions. The TH approach is quadratic
in $\sigma$ as well as $\mid\gamma_{T}\mid T_{h}$ (Eq. \ref{eq:Thermo-Disp-TC}),
while, in the LSA case (Eq. \ref{eq:LSA-Disp-TC}) it is linear in
both quantities. Consequently, the TH approach does not directly lead
to the linear dependence on $\mid\gamma_{T}\mid T_{h}$ as evident
from LSA (Eq. \ref{eq:LSA-Disp-TC}). This is especially important
since, as noted earlier, the thermal gradients generated by pulsed
laser heating can have positive or negative signs and hence lead to
fundamentally different dewetting behaviors \cite{KrishnaPCCP09,trice08}.
Since the TH dispersion is a quadratic function of the thermal gradient,
its behavior will be independent of the sign of the thermal gradient
and so does not agree with LSA. As we show next, it is necessary to
evaluate the characteristics of dissipation in order to get the correct
behavior from TH.

\subsubsection{Dispersion using minimum viscous dissipation}

In the classical case, the total viscous dissipation is uniquely defined
by the magnitude of the pressure gradient for any given film thickness,
as evident from Eq. \ref{eq:Evtotal_Class}. On the other hand, the
total dissipation for thermocapillary dewetting is not unique, and,
in fact, varies with the magnitude of the thermal gradient for any
given pressure gradient, as evident from Eq. \ref{eq:Evtotal_TC}.
It is this behavior that is responsible for the above discrepancy
between LSA and TH and can be resolved by evaluating the minimum viscous
dissipation. 

The minimum viscous dissipation for the fluid being subjected to pressure
gradients can be estimated from the differential condition $d\dot{e}/d\nabla P=0$.
Using Eq. \ref{eq:ev_TC}, this leads to the condition $\nabla P(z-h_{0})=\mid\gamma_{T}\mid T_{h}h'$.
The general solutions satisfying the above equality can be evaluated
for various values of height $z$ in relation to the thickness $h_{o}$.
First, the condition $z=h_{0}$ does not yield a unique relation between
$\nabla P$ and $\mid\gamma_{h}\mid h'$ and is therefore not a useful
solution in the context of the dissipation. On the other hand, the
choice of $z=0$, yields the case of $\nabla P=-\frac{\mid\gamma_{T}\mid T_{h}h'}{h_{0}}$.
By utilizing Eq. \ref{eq:dVxoverdZ}, one can see that the physical
interpretation of this condition is that the tangential stress at
the film-substrate interface at $z=0$ is zero. The resulting viscous
dissipation for thermocapillary dewetting will now be (from Eq. \ref{eq:ev_TC}):\begin{equation}
\dot{e}_{TC}^{m}=\frac{(\nabla Pz)^{2}}{\eta}\label{eq:Viscous-dissi}\end{equation}
where the superscript \emph{m} signifies a minimum. One can verify
that this is a minima by noting that the second derivative $d^{2}\dot{e}_{v}/d^{2}\nabla P$
is positive. Therefore, the minimum dissipation per unit area of the
film $\dot{E}_{TC}^{m}$ can be obtained as: \begin{equation}
\dot{E}_{TC}^{m}=\intop_{o}^{h_{o}}\dot{e}_{TC}^{m}dz=\frac{(\nabla P)^{2}h_{0}^{3}}{3\eta}\label{eq:Min-viscous-dissi}\end{equation}

\noindent Using the form of $\nabla P$ from Eq. \ref{eq:nablaPfinal_TC},
the relevant form of $\dot{E}_{TC}^{m}$ is:

\noindent \begin{equation}
\dot{E}_{TC}^{m}=-\left(\frac{3\eta}{h_{0}^{3}k^{2}}\sigma^{2}+\frac{3\mid\gamma_{h}\mid}{h_{0}}\sigma+\frac{3(\mid\gamma_{h}\mid)^{2}k^{2}h_{0}}{4\eta}\right)\epsilon^{2}e^{2\left(\sigma t-ikx\right)}\label{eq:Evtotal_TC_Min}\end{equation}

\noindent Finally, by equating the rate of free energy change $\dot{\Delta F}$
(Eq. \ref{eq:dFoverdt}) and the minimum viscous dissipation rate
$\dot{E}_{TC}^{m}$ (Eq. \ref{eq:Evtotal_TC_Min}), we obtain an analytical
form of the dispersion relation as: \begin{equation}
\sigma^{2}+\frac{h_{0}^{3}k^{2}}{3\eta}(\gamma k^{2}+\frac{A}{2\pi h_{0}^{4}}+\frac{3\mid\gamma_{T}\mid T_{h}}{h_{o}})\sigma+\frac{h_{0}^{3}k^{2}}{3\eta}\frac{3}{4\eta}(\mid\gamma_{T}\mid T_{h})^{2}h_{0}k^{2}=0\label{eq:dispersion relation}\end{equation}

\noindent This result is clearly different from Eq. \ref{eq:Thermo-Disp-TC}
because now, the linear behavior with $\mid\gamma_{T}\mid T_{h}$
is also present. Next, we evaluate this quadratic equation in $\sigma$
for various magnitudes of the thermal gradients and show that it is
identical to the LSA results for large thermal gradients. By defining
$f=\gamma k^{2}+A/2\pi h_{0}^{4}$ and $g=(3\mid\gamma_{T}\mid T_{h})/h_{0}$,
the roots of the dispersion relation are: \begin{equation}
\sigma_{\pm}=-\frac{h_{0}^{3}k^{2}}{6\eta}(f+g)\pm\frac{h_{0}^{3}k^{2}}{6\eta}\sqrt{(f+g)^{2}-g^{2}}\label{eq:Thermo-DispRts-TC-Min}\end{equation}

\begin{enumerate}
\item Minimum dissipation approach for classical case (i.e. $T_{h}=0)$

By substituting $T_{h}=g=0$ in Eq. \ref{eq:Thermo-DispRts-TC-Min},
the resulting relevant root is:

\noindent \begin{equation}
\sigma_{-}=-\frac{h_{0}^{3}k^{2}}{3\eta}(\gamma k^{2}+\frac{A}{2\pi h_{0}^{4}})\label{eq:dispersion-noTh}\end{equation}

As expected, this result is identical to the result for the classical
dewetting instability. 

\item Minimum dissipation approach for strong thermal gradients

\noindent In the case when the magnitudes of the thermal gradients
are larger then the attractive dispersion forces, i.e. for example
when $3\mid\gamma_{T}T_{h}\mid\geq\mid A/2\pi h_{o}^{3}\mid$, then
we have $\mid g\mid>\mid f\mid$. This is the condition found in the
experimental cases reported earlier \cite{KrishnaPCCP09,trice08},
and we can obtain an approximate solution from Eq. \ref{eq:Thermo-DispRts-TC-Min}
as follows:\begin{equation}
\sigma_{\pm}\approxeq-\frac{h_{o}^{3}k^{2}}{6\eta}(f+g)\pm\frac{h_{o}^{3}k^{2}}{6\eta}\sqrt{g^{2}-g^{2}}=-\frac{h_{o}^{3}k^{2}}{6\eta}(f+g)\label{eq:dissipation strongTh}\end{equation}

or

\begin{equation}
\sigma_{TC}^{TH}=-\frac{h_{o}^{3}k^{2}}{3\eta}(\gamma k^{2}+\frac{A}{2\pi h_{o}^{4}}+\frac{3\mid\gamma_{T}\mid T_{h}}{h_{o}})\label{eq:Thermo-DispRt-TC}\end{equation}

\noindent From Eq. \ref{eq:Thermo-DispRt-TC}, the characteristic
wavelength $\Lambda_{TC}^{TH}$ obtained from the maxima in the dispersion
given by $d\sigma/dk=0$ is: 

\noindent \begin{equation}
\Lambda_{TC}^{TH}=\sqrt{-\frac{16\pi^{3}\gamma}{A+6\pi\mid\gamma_{T}\mid T_{h}h_{o}^{3}}}h_{o}^{2}\label{eq:ThermoLambdaTC}\end{equation}

Comparing with the LSA result, Eq. \ref{eq:LSA_LambdaTC} , the only
difference is in the factor of two multiplying the thermal gradient
term. More importantly, the functional dependence on $h_{o}$, A,
$\gamma$ and $\mid\gamma_{T}\mid T_{h}$ remains the same, confirming
that the minimum dissipation approach gives similar physical characteristics
for the dewetting instability. An important benefit of utilizing the
thermodynamic approach is also evident here. From the above dissipation
analysis it is clear that there are multiple choices for the dewetting
pathway in regards to the rate of dissipation. However the instability
clearly picks the path which minimizes the rate of this dissipation,
or in other words, minimizes the overall rate at which energy is lost
in the dewetting process. We have also plotted the behavior of the
dispersion relation, Eq. \ref{eq:Thermo-DispRt-TC} for Co films on
SiO$_{\text{2}}$ substrates with the various materials parameter
values being: Hamaker coefficient A = $-1.41\times10^{-18}$~J, $\gamma=1.88$~J/m$^{\text{2}}$
and $\gamma_{T}$ = $-4.8\times10^{-3}\, J/m^{2}-K$. Fig. \ref{fig:Dispersion}(a)\textit{\emph{
plots the growth rate $\sigma$ versus wave number k for the dispersion
relation }}Eq. \ref{eq:Thermo-DispRt-TC}\textit{\emph{. The data
was evaluated for Co film of thickness 8 nm and various values of
the gradient $T_{h}$, }}including $T_{h}=0$, $T_{h}>0$ and $T_{h}<0$\textit{\emph{,
as indicated on the plot. In Fig. \ref{fig:Dispersion}(b) ) we have
plotted the characteristic length scale }}\emph{$\Lambda_{TC}^{TH}$}\textit{\emph{
for Co films on SiO$_{\text{2}}$ as a function of varying thickness
and various $T_{h}$.}} The decrease in length scale with increase
in the magnitude of $T_{h}<0$ is evident. In Fig. \ref{fig:Dispersion}(c)
the\textit{\emph{ cut-off wavelength}}$\Lambda_{cut-off}^{TH}$ \textit{\emph{as
a function of varying thickness and various $T_{h}$ is plotted. The
cut-off wavelength corresponds to the intersection of the growth rate
with the wave number axis in Fig. \ref{fig:Dispersion}(a).}}

\item Minimum dissipation approach for weak thermal gradients

\noindent In the case when the magnitudes of the thermal gradients
are smaller then the attractive dispersion forces, i.e. for example
when $3\mid\gamma_{T}T_{h}\mid<\mid A/2\pi h_{o}^{3}\mid$, then we
have $\mid g\mid<\mid f\mid$. In this situation, we can approximate
Eq. \ref{eq:Thermo-DispRts-TC-Min} as follows: \begin{equation}
\sigma_{\pm}\approxeq-\frac{h_{o}^{3}k^{2}}{6\eta}(f+g)\pm\frac{h_{o}^{3}k^{2}}{6\eta}\sqrt{f^{2}-g^{2}}\approxeq-\frac{h_{o}^{3}k^{2}}{6\eta}(f+g)\pm\frac{h_{o}^{3}k^{2}}{6\eta}f(1-\frac{g^{2}}{2f^{2}})\end{equation}

\end{enumerate}
\noindent where we have used the binomial approximation $(1-(g/f)^{2})^{1/2}\sim(1-g^{2}/2f^{2})$.
Here we find again that the dispersion is considerably different from
the LSA result of Eq. \ref{eq:LSA-Disp-TC}.

\section{Conclusion}

We have theoretically evaluated the classical and thermocapillary
dewetting instability in thin fluid films via a thermodynamic approach.
In this, the rate of change of free energy is equated to the viscous
dissipation in the thin film. The thermodynamic approach leads to
an analytical expression for the dispersion without the need for explict
solutionof the height evolution dynamical equation from NS within
the lubrication approximation. We have compared results from this
approach to existing results obtained by linearization of the fluid
dynamics of the thin film. For the case of classical dewetting in
the presence of surface tension and long range attractive forces,
the thermodynamic approach predicts identical behavior to that from
linear analysis. We have also evaluated dewetting in the presence
of film-thickness dependent temperature variations. Such a situation
can be found during dewetting of thin metallic films melted by a nanosecond
ultraviolet pulsed laser. In this condition, a film thickness dependent
reflection and absorption leads to thermocapillary forces along the
plane of the film. In this scenario we found that the thermodynamic
approach agrees with linear analysis provided the minimum viscous
dissipation is evaluated. The fluid flow condition that gives minimum
viscous dissipation is one where the film-substrate tangential stress
is zero. In the context of dewetting in the presence of film thickness
dependent thermocapillary forces, the thermodynamic approach clearly
illustrates that the instability chooses a pathway which minimizes
the rate of energy loss in the system. This results shows that the
thermodynamic approach based on evaluating the rates of free energy
change and energy loss is a simple but potentially powerful way to
gain physically meaningful insight into such spontaneous pattern formation
processes.

RK acknowledges support by the National Science Foundation through
CAREER grant NSF-DMI-0449258, grant NSF-CMMI-0855949, and grant NSF-DMR-0856707.

\bibliographystyle{ieeetr}

\pagebreak

\section*{Figure captions}
\begin{figure}[!tbh]
\begin{raggedright}
\subfloat[]{\begin{raggedright}
\includegraphics[clip,width=4in]{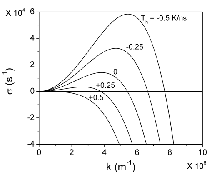}
\par\end{raggedright}

}
\par\end{raggedright}

\begin{raggedright}
\subfloat[]{\begin{raggedright}
\includegraphics[clip,width=4in]{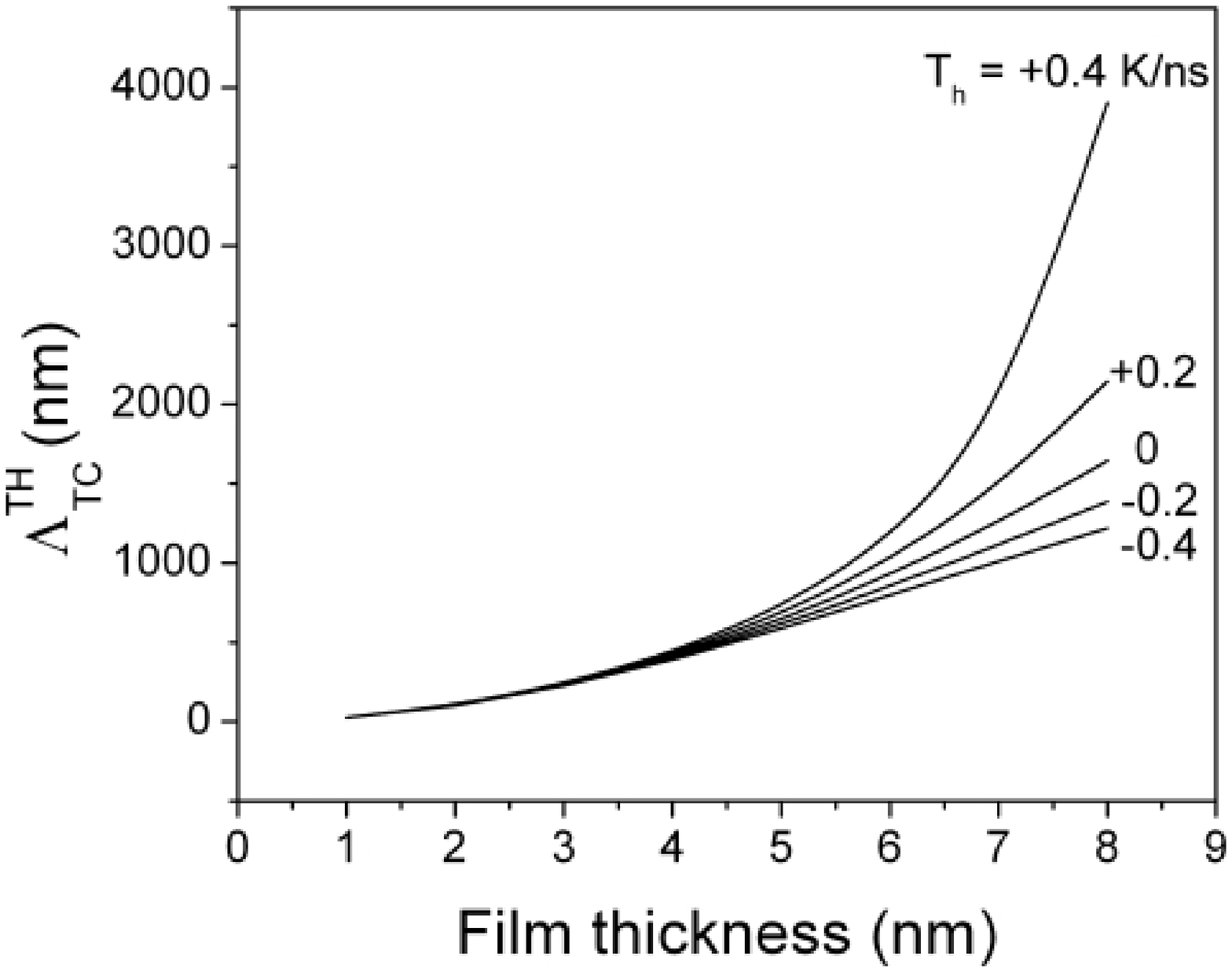}
\par\end{raggedright}

}
\par\end{raggedright}

\begin{raggedright}
\subfloat[]{\begin{raggedright}
\includegraphics[clip,width=4in]{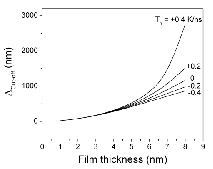}
\par\end{raggedright}

}
\par\end{raggedright}

\caption{\textit{(a) Plot of the growth rate versus wave number for the dispersion
relation obtained from the dissipation approach for strong thermal
gradients. The data was evaluated for Co film of thickness 8 nm and
various values of the gradient $T_{h}$, as indicated on the plot.
(b) Plot of the characteristic length scale }$\Lambda_{TC}^{TH}$\textit{
for Co films on SiO$_{\text{2}}$ as a function of various $T_{h}$.
(c) Plot of the cut-off wavelength}$\Lambda_{Cutt-off}$\textit{ for
Co films on SiO$_{\text{2}}$ as a function of various $T_{h}$. The
cut-off wave number corresponds to the intersection of the rate with
the wave number axis in Fig. (a).}\label{fig:Dispersion}}

\end{figure}

\begin{itemize}
\item \textbf{Figure \ref{fig:Dispersion}}: \textit{(a) Plot of the growth
rate versus wave number for the dispersion relation obtained from
the dissipation approach for strong thermal gradients. The data was
evaluated for Co film of thickness 8 nm and various values of the
gradient $T_{h}$, as indicated on the plot. (b) Plot of the characteristic
length scale }$\Lambda_{TC}^{TH}$\textit{ for Co films on SiO$_{\text{2}}$
as a function of various $T_{h}$. (c) Plot of the cut-off wavelength}$\Lambda_{Cutt-off}$\textit{
for Co films on SiO$_{\text{2}}$ as a function of various $T_{h}$.
The cut-off wave number corresponds to the intersection of the rate
with the wave number axis in Fig. (a).}
\end{itemize}
\pagebreak

\end{document}